\documentclass[preprint,showpacs]{revtex4}

\usepackage{graphicx}
\usepackage{bm}

\begin{document}

\title{Spin observables in nucleon-deuteron scattering and three-nucleon forces}
\author{S. Ishikawa} \email[E-mail:]{ishikawa@i.hosei.ac.jp}
\author{M. Tanifuji}
\affiliation{Department of Physics, Science Research Center, Hosei University, 
Fujimi 2-17-1, Chiyoda, Tokyo 102-8160, Japan}
\author{Y. Iseri}
\affiliation{Department of Physics, Chiba-Keizai College, 
Todoroki-cho 4-3-30, Inage, Chiba 263-0021, Japan}

\date{\today}

\begin{abstract}
Three-nucleon forces, which compose an up-to-date subject in few-nucleon systems, provide a good account of the triton binding energy and the cross section 
minimum in proton-deuteron elastic scattering, while do not succeed in explaining spin observables such as the nucleon and deuteron analyzing powers, suggesting serious defects in their spin dependence. 
We study the spin structure of nucleon-deuteron elastic amplitudes by decomposing them into spin-space tensors and examine effects of three-nucleon forces
to each component of the amplitudes obtained by solving the Faddeev equation.
Assuming that the spin-scalar amplitudes dominate the others, we derive simple expressions for spin observables in the nucleon-deuteron elastic scattering. 
The expressions suggest that a particular combination of spin observables in the scattering provides direct information of scalar, vector, or tensor component of the three-nucleon forces.
These effects are numerically investigated by the Faddeev calculation.
\end{abstract}

%

\maketitle

\section{Introduction}
\label{Sec_1}

In the last decade, a lot of investigations have been devoted to nucleon-deuteron ($Nd$) elastic scattering to provide valuable information on nuclear interactions such as an evidence of a three-nucleon force (3NF) \cite{Gl96,Sa94,Sh95,Sa96,St99,Bi00,Sa00,Ca01,Br01,Er01,Wo02,Se02,Mc94,Is99,Wi01,Ki01}. 
In these studies, serious discrepancies between theoretical predictions with conventional models of two-nucleon force (2NF) and corresponding experimental data were found for some scattering observables in addition to the well known underbinding problem of the three-nucleon (3N) bound states. 
An example of the discrepancies is that calculated nucleon- and deuteron vector analyzing powers are considerably smaller than the measured in a low energy region, say $E_p \alt 20$ MeV \cite{Ko87,Wi88,Sa94,Sh95,Br01,Wo02,Mc94,Ki01}. 
Another example of the discrepancies is observed in the proton-deuteron ($pd$) elastic cross sections around the minima of angular distributions, where the 2NF calculations  underestimate the cross section systematically when compared to the measured ones at intermediate energies $E_p=60 - 150$ MeV \cite{Sa96,Sa00,Se02}.
The latter discrepancy has been solved by introducing a two-pion exchange three-nucleon force ($2\pi$E-3NF) with a cutoff parameter adjusted so as to reproduce the empirical binding energy of the triton \cite{Wi98}.
However, the introduction of the 3NF provides only a small effect on the former discrepancy, or sometimes gives rise to worse agreements with experimental data on some spin observables \cite{St99,Bi00,Sa00,Ca01,Er01,Se02,Wi01}. 
We are, therefore, still far from the final understanding of nuclear interactions in the 3N system.  

Since the difficulties are concerned with the spin observables, detailed investigations on the contribution of spin-dependent interactions in the observables will be required for improving the relevant interactions. 
Previously \cite{Is01} we have analyzed the imaginary parts of neutron-deuteron ($nd$) forward scattering amplitudes, which are transformed into combinations of total cross sections for polarized or unpolarized beams and targets by the use of the optical theorem. 
By decomposing the forward amplitudes according to their spin-space properties and examining the contributions of spin-dependent interactions to the components, we have succeeded in clarifying spin-dependent effects of the 3NF on combinations of the total cross sections. 
Being encouraged by such success, we will here develop the previous work to the case of $Nd$ spin observables at finite scattering angles. 
As in Ref.\ \cite{Is01}, we will examine the contribution of a particular spin-dependent interaction on the observables by tagging ranks of related spin-space tensors on the scattering amplitude as described below, and then will study which observable characterizes the effect of the particular spin-dependent interaction. 

For the tagging, we will decompose scattering amplitudes according to the tensorial property in the spin-space. 
In such decomposition, where we obtain scalar amplitude, vector one, second-rank tensor one, and so on, each component specified by the tensor rank will describe the scattering by corresponding interactions: 
the scalar component will describe the scattering by central interactions, the vector one the scattering by spin vector interactions like spin-orbit ones, and the second-rank tensor one the scattering by tensor interactions. 
In the next section, such decomposition of the scattering amplitude is performed  in a model-independent way by the invariant amplitude method \cite{Ta68}. 

When the observables are described in terms of such decomposed amplitudes, one will be able to identify the contribution of the particular spin-dependent interaction by the help of the associated tensor rank. 
However, full expressions of the observables in terms of the decomposed scattering amplitudes are rather complicated. 
In Sec.\ \ref{Sec_3}, we will show that the central interaction will dominate the scattering at low energies. 
Then the observables are described rather simply in an approximation neglecting second order terms of the vector and tensor amplitudes in the expressions. 
This approximation gives a clear insight into the role of each component of the interaction and one can obtain the information on the observables, which characterize the contribution of the component. 
Utilizing the results, typical observables such as the vector and tensor analyzing powers and spin correlation coefficients are fully analyzed by the Faddeev calculations in Sec.\ \ref{Sec_4}. 
In the analyses, investigations are particularly focused on the contributions of 3NFs to the observables, which include various spin effects and are decomposed into scalar effects, vector ones, and tensor ones. 
The incident energy is fixed to $E_N=3$ MeV, since at low energies below the deuteron breakup threshold, the Coulomb interaction can be treated exactly in the Faddeev calculation \cite{Is02}. 

Matrix elements of a scattering T-matrix in terms of the invariant amplitudes are given in the Appendix \ref{App_A}, and formulae 
of polarization transfer coefficients in the approximation are given in the Appendix \ref{App_B} for the convenience of applications of the theory.


\section{T-matrix and transition amplitudes}
\label{Sec_2}

Let us describe a scattering T-matrix $\bm{M}$ for the $Nd$ scattering by specifying the elements by the $z$ component of the deuteron spin $\nu_d$ and that of the nucleon spin $\nu_N$ as
\begin{equation}
\bm{M}
=\left(
\begin{array}{cccccc}
 A & B & C & D & E & F \\
 G & H & I & J & K & L \\
 M & N & O & P & Q & R \\
 R &-Q &-P & O & N &-M \\
 L &-K &-J & I & H &-G \\
-F & E & D &-C &-B & A
\end{array}
 \right),
\label{eq1}
\end{equation} 
where the row and column are designated by 
($\nu_d=1, \nu_N=\frac12$), ($\nu_d=1, \nu_N=-\frac12$), 
($\nu_d=0, \nu_N=\frac12$), ($\nu_d=0, \nu_N=-\frac12$), 
($\nu_d=-1, \nu_N=\frac12$), ($\nu_d=-1, \nu_N=-\frac12$) 
from left to right for the initial state and from top to bottom for the final state. 
These matrix elements will explicitly be described by scattering amplitudes that have particular tensorial property in the spin space. 
For that purpose, we will decompose $\bm{M}$ by spin-space tensors $\bm{S}_{K\kappa}$, where $K$ and $\kappa$ are the rank of the tensor and its $z$ component,
\begin{equation}
 \bm{M} = \sum_K \bm{M}_K, \quad 
 \bm{M}_K = \sum_{\kappa} (-)^{\kappa} \bm{S}_{K-\kappa} \bm{R}_{K\kappa}.
\label{eq2}
\end{equation}   
Here, $\bm{R}_{K\kappa}$ is the counter part, a tensor in the coordinate space.  Using Eq.\ (\ref{eq2}), matrix elements of $\bm{M}_K$ are given \cite{Ta68} by 
\begin{eqnarray}
&&<\nu_N^\prime \nu_d^\prime; \bm{k}_f| \bm{M}_K |\nu_N \nu_d; \bm{k}_i> 
\nonumber \\
 &=& \sum_{s_is_f}(s_N s_d\nu_N\nu_d|s_i\nu_i)
     (s_N s_d \nu_N^{\prime} \nu_d^{\prime}|s_f \nu_f)(-)^{s_f-\nu_f} 
\nonumber\\ 
 & & \times (s_is_f \nu_i -\nu_f|K\kappa) \sum_{\ell_i=\bar{K}-K}^K
  \bigl[C_{\ell_i}(\hat{k}_i) \otimes C_{\ell_f}(\hat{k}_f)\bigr]_{\kappa}^K 
\nonumber\\ 
& & \times F(s_is_fK \ell_i),
\label{eq3}
\end{eqnarray} 
where $s$ is the spin of the related particle, 
$\bm{k}_i (\bm{k}_f)$ is the $Nd$ relative momentum in the initial (final) state, 
$C_{\ell m}(\Omega)=\bigl(4\pi/(2\ell +1)\bigr)^{\frac12} Y_{\ell m}(\Omega)$, 
and $\bar{K}=K$ for even $K$ and $\bar{K}=K+1$ for odd $K$.  
The amplitude $F(s_is_fK \ell_i)$ is called the invariant amplitude due to the invariance under rotations of the coordinate axes and is a function of the center of mass energy and the scattering angle $\theta$. 
In Eq.\ (\ref{eq3}), while the geometrical part of the matrix element of $\bm{S}_{K-\kappa}$ is described by the Clebsch-Gordan coefficients using the Wigner-Eckart theorem and that of $\bm{R}_{K\kappa}$ is represented by 
$\bigl[C_{\ell_i}(\hat{k}_i) \otimes C_{\ell_f}(\hat{k}_f)\bigr]_{\kappa}^K$, their physical parts are included in $F(s_is_fK \ell_i)$, which represents the scattering by spin-space scalar interactions for $K=0$, the one by vector interaction for $K=1$, and so on.

For the $Nd$ scattering, in which $s_i$ and $s_f$ are $\frac12$ or $\frac32$, $K$ takes $0$, $1$, $2$ and $3$. 
For the components of the amplitude in Eq.\ (\ref{eq3}), we will denote 
the scalar amplitudes ($K=0$), the vector ones ($K=1$), the tensor ones ($K=2$), and the third-rank tensor ones ($K=3$) as follows:
\begin{equation}
\left\{ \begin{array}{ccl}
U_1 &\equiv& F(\frac12 \frac12 0 0)
\\ \\
U_3 &\equiv& F(\frac32 \frac32 0 0)
\end{array}\right.
\label{eq4}
\end{equation}
\begin{equation}
\left\{ \begin{array}{ccl}
S_1 &\equiv& \bigl[C_1(\hat{k}_i) \otimes C_1(\hat{k}_f)\bigr]_1^1 F(\frac12 \frac12 11)
\\ \\
S_2 &\equiv& \bigl[C_1(\hat{k}_i) \otimes C_1(\hat{k}_f)\bigr]_1^1 F(\frac32 \frac12 11)
\\ \\
S_3 &\equiv& \bigl[C_1(\hat{k}_i) \otimes C_1(\hat{k}_f)\bigr]_1^1 F(\frac32 \frac32 1 1)
\\ \\
S_4 &\equiv& \bigl[C_1(\hat{k}_i) \otimes C_1(\hat{k}_f)\bigr]_1^1 F(\frac12 \frac32 11)
\end{array}\right.
\label{eq5}
\end{equation}
\begin{equation}
\left\{ \begin{array}{ccl}
T_1(\kappa) &\equiv& \displaystyle{\sum_{\ell_i} \bigl[C_{\ell_i} (\hat{k}_i) \otimes C_{\ell_f} (\hat{k}_f) \bigr]_{\kappa}^2 F(\frac32 \frac12 2 \ell_i)}
\\
T_2(\kappa) &\equiv& \displaystyle{\sum_{\ell_i} \bigl[C_{\ell_i} (\hat{k}_i) \otimes C_{\ell_f} (\hat{k}_f) \bigr]_{\kappa}^2 F(\frac12 \frac32 2 \ell_i)}
\\
T_3(\kappa) &\equiv& \displaystyle{\sum_{\ell_i} \bigl[C_{\ell_i} (\hat{k}_i) \otimes C_{\ell_f} (\hat{k}_f) \bigr]_{\kappa}^2 F(\frac32 \frac32 2\ell_i)}
\\
&& \qquad\qquad (\kappa=0, 1, 2) 
\end{array}\right.
\label{eq9}
\end{equation}
\begin{eqnarray}
V(\kappa) &\equiv& \displaystyle{\sum_{\ell_i} 
 \bigl[C_{\ell_i} (\hat{k}_i) \otimes C_{\ell_f} (\hat{k}_f) \bigr]_{\kappa}^3 
 F(\frac32 \frac32 3 \ell_i)}
\nonumber \\
 && \qquad \qquad (\kappa=1, 2, 3)
\label{eq10}
\end{eqnarray}
Here, $U_1$ and $S_1$ describe the scattering in the spin doublet state $(s_i=s_f=\frac12)$, $U_3$, $S_3$, $T_3(\kappa)$, and $V(\kappa)$ describe those in the spin quartet state $(s_i=s_f=\frac32)$, and $S_2$, $S_4$, $T_1(\kappa)$, and $T_2(\kappa)$ describe the doublet-quartet nondiagonal transitions. 
The time reversal theorem gives one relation,
\begin{equation}
S_4=-S_2,
\label{eq15}
\end{equation}
for the vector amplitudes, four relations between the nine tensor amplitudes, and one relation between the three third-rank tensor amplitudes, although latter five relations are not used explicitly. 
These relations are equivalent to those given for $d$+${}^3$He scattering in Ref.\ \cite{Je80}.

The matrix elements $A$, ..., $R$ are described in terms of the amplitudes $U_1$, ..., $V(3)$, whose explicit expressions are given in Appendix \ref{App_A}.
Solving Eqs.\ (\ref{eq16})-(\ref{eq31}) in Appendix \ref{App_A} inversely,  we get the amplitudes $U_1$, ..., $V(3)$ in terms of $A$, ..., $R$ as follows:
\begin{eqnarray}
U_1 &=&\frac{2\sqrt2}3 H-\frac23 I -\frac23 N +\frac{\sqrt2}3 O
\nonumber\\
U_3 &=& A+\frac13 H +\frac{\sqrt2}3 I +\frac{\sqrt2}3 N +\frac23 O
\label{eqA1}
\end{eqnarray}
\begin{eqnarray}
S_1&=&\frac13 \bigl(\sqrt2 J-2K-P+\sqrt2 Q\bigr) 
\nonumber\\
S_2&=&-\frac16\bigl(3\sqrt2 G-3M+2J+\sqrt2 K-\sqrt2 P -Q\bigr)
\nonumber\\
S_3&=&\frac1{3\sqrt{10}} \bigl(3B +3\sqrt2 C-3G-3\sqrt2 M +2\sqrt2 J 
\nonumber\\
&&+2K +4P +2\sqrt2 Q\bigr)
\nonumber\\
S_4&=&-\frac16\bigl(3\sqrt2 B -3C +J -\sqrt2 K 
\nonumber\\
&&+\sqrt2 P -2Q\bigr)
\label{eqA5}
\end{eqnarray}
\begin{eqnarray}
T_1(0)&=&-\frac23 H-\frac{2\sqrt2}3 I +\frac{\sqrt2}3 N +\frac23 O
\nonumber\\
T_2(0)&=&\frac23 H -\frac{\sqrt2}3 I +\frac{2\sqrt2}3 N -\frac23 O
\nonumber\\
T_3(0)&=&A-\frac13 H-\frac{\sqrt2}3 I-\frac{\sqrt2}3 N-\frac23 O
\label{eqA2}
\end{eqnarray}
\begin{eqnarray}
T_1(1)&=&-\frac1{2\sqrt3}\bigl(\sqrt2 G-M-2J-\sqrt2 K +\sqrt2 P +Q\bigr)
\nonumber\\
T_2(1)&=&\frac1{2\sqrt3}\bigl(\sqrt2 B -C-J+\sqrt2 K-\sqrt2 P +2Q\bigr)
\nonumber\\
T_3(1)&=&-\frac1{\sqrt6} \bigl(B+\sqrt2 C+G+\sqrt2 M\bigr)
\label{eqA3}
\end{eqnarray}
\begin{eqnarray}
T_1(2)&=&-\frac1{\sqrt3}\bigl(\sqrt2 L-R\bigr)
\nonumber\\
T_2(2)&=&-\frac1{\sqrt3} \bigl(D-\sqrt2 E\bigr) 
\nonumber\\
T_3(2)&=&\frac1{\sqrt6}\bigl(\sqrt2 D+E+L+\sqrt2 R\bigr)
\label{eqA4}
\end{eqnarray}
\begin{eqnarray}
V(1)&=&\frac1{\sqrt{15}} \bigl(B+\sqrt2 C-G -\sqrt2 M-\sqrt2 J-K
\nonumber\\
&& -2P-\sqrt2 Q\bigr)
\nonumber\\
V(2)&=&\frac1{\sqrt6} \bigl(L+\sqrt2 R-\sqrt2 D -E\bigr)
\nonumber\\
V(3)&=&F
\label{eqA6}
\end{eqnarray}

Intrinsic third-rank tensor interactions are unknown and possible third-rank tensor amplitudes may arise from higher orders of the vector and tensor interactions, which is supposed to have small contributions to the scattering. 
Then the third-rank tensor amplitudes will be neglected in later applications for simplicity. 

When effective interactions are introduced by a $Nd$ two-body model, one can directly relate the amplitude $U_j$, $T_j(\kappa=0,1,2)$, and $S_j$ to the components of the model interactions, i.e., central ones, tensor ones, and spin-orbit ones.
The example of the relation is given in Ref.\ \cite{Is01}.
Then the analyses in the following sections can be represented in terms of such effective interactions when necessary.

\section{Analyses of Invariant  Amplitudes}
\label{Sec_3}

\begin{figure*}[t]
\includegraphics[scale=0.5]{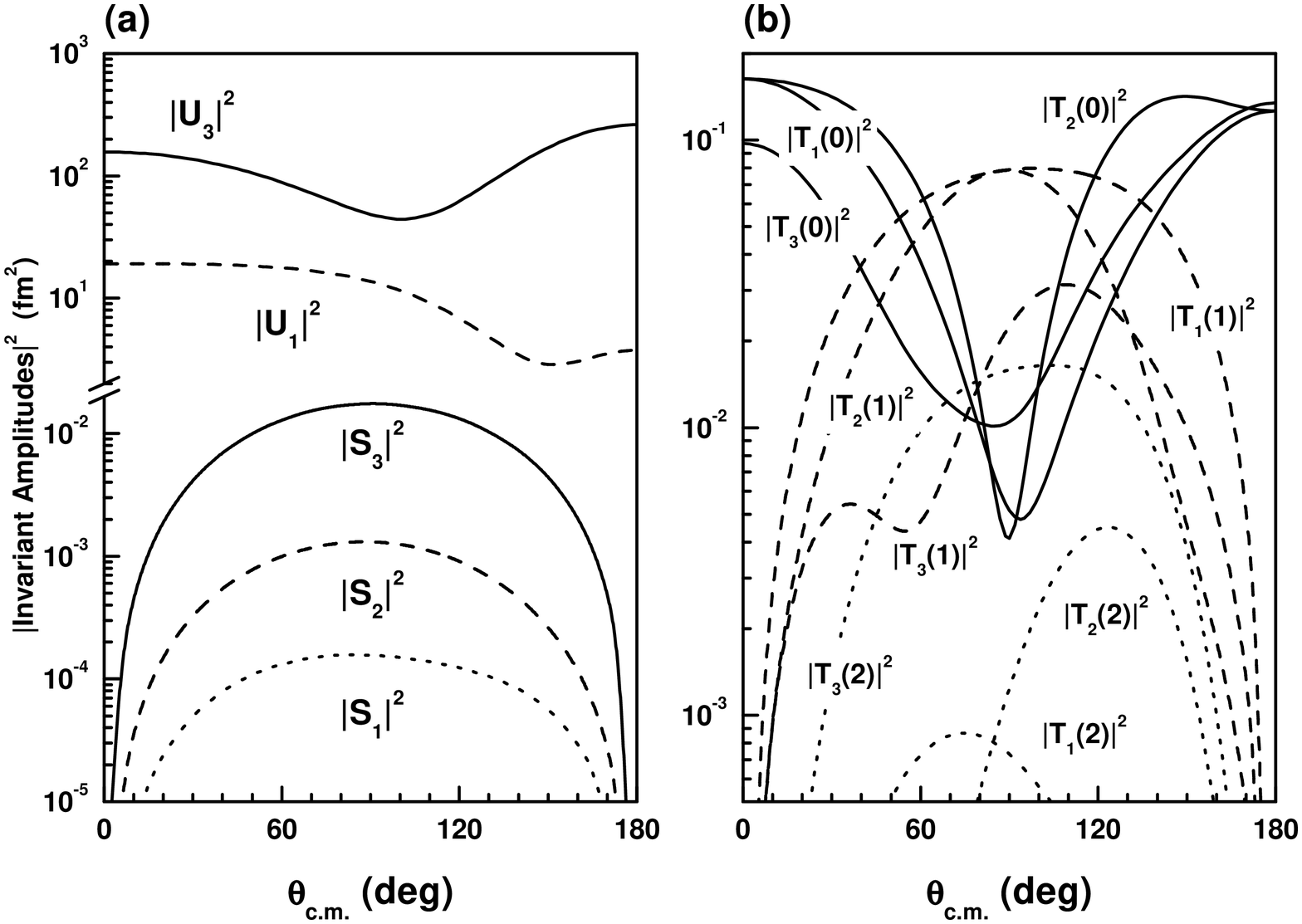}
\caption{\label{fig1}
The absolute squares of the scattering amplitudes $U_1$, $U_3$, $S_1$, $S_2$, and $S_3$ are shown in (a), and those of $T_1(\kappa)$, $T_2(\kappa)$, and $T_3(\kappa)$ for $\kappa=0, 1, 2$ are shown in (b). 
The curves are the Faddeev calculations with the AV18 for the $nd$ scattering at $E_n=3$ MeV. 
}
\end{figure*}

In the present work, we have calculated the scattering amplitudes by solving the Faddeev equation in the coordinate space \cite{Is87} with the Argonne V$_{18}$ model (AV18) \cite{Wi95} for the input 2NF. 
The Coulomb interaction is included for the $pd$ scattering by the method in Ref.\ \cite{Is02}. 
The 3N partial wave states for which the 2NF acts are restricted to those with total two-nucleon angular momenta up to 2, and the total 3N angular momentum is truncated at $19/2$. 
The accuracy of our calculations is examined by the comparison with variational calculations for the AV18 2NF with the pair correlated hyperspherical harmonics basis \cite{Ki95}, whose results of the AV14 2NF \cite{Wi84} are considered as benchmarks for the $nd$ scattering \cite{Hu95} and the $pd$ scattering \cite{Ki01b}.
The agreement of the phase-shift parameters by both methods 
are found within a few percent deviation for all of thirty phase-shift parameters up to the $5/2^{-}$ state \cite{Is02}.

Calculated binding energy of the triton for the AV18 is 7.51 MeV, which is small compared to the empirical value of 8.48 MeV. 
An additional contribution of the 2$\pi$E-3NF is investigated by using the Brazil model (BR) \cite{Co83}, which gives 8.44 MeV for the triton binding energy in combination with AV18 (AV18+BR).  
Furthermore, we introduce two kinds of the phenomenological models of the 3NF.  
One is for the study of the role of the central force part of the BR-3NF, which is simulated by a Gaussian type (GS) \cite{Is99} as
\begin{equation}
V_{\text{GS-3NF}} = 
  V_0^G \sum_{i \ne j \ne k}
  \exp\left\{ -(\frac{r_{ji}}{r_G})^2-(\frac{r_{ki}}{r_G})^2 \right\}. 
\label{41}
\end{equation}
Values of the parameters, which are determined so as to reproduce 
the empirical triton binding energy of 8.48 MeV in combination with AV18 (AV18+GS), are $r_G=1.0$ fm and $V_0^G=-45$ MeV. 
The other model of 3NF is the spin-orbit (SO) 3NF \cite{Ki99}, which is adopted as an example of new spin vector interactions to account for the discrepancy in the vector analyzing powers, 
\begin{equation}
V_{\text{SO-3NF}} = \frac12 W_0 \exp\{ -\alpha \rho\}  \sum_{i < j} 
  \left(\bm{l}_{ij} \cdot ( \bm{\sigma}_i + \bm{\sigma}_j) \right) 
  {\hat P}_{11},
\label{eq42} 
\end{equation}
with $\rho^2=\frac23 \bigl(r_{12}^2 +r_{23}^2 +r_{31}^2\bigr)$ and ${\hat P}_{11}$ is the projection operator to the spin and isospin triplet state of the pair $(i,j)$. 
For this interaction several parameter sets are suggested in Ref.\ \cite{Ki99}, among which we take $\alpha =$ 1.5 fm$^{-1}$ and $W_0 =$ -20 MeV. 
The SO-3NF gives a repulsive effect on the triton binding energy. 
The resultant binding energy for the SO-3NF with the BR-3NF (AV18+BR+SO) is 8.39 MeV.

%
\begin{figure*}[tbh]
\includegraphics[scale=0.5]{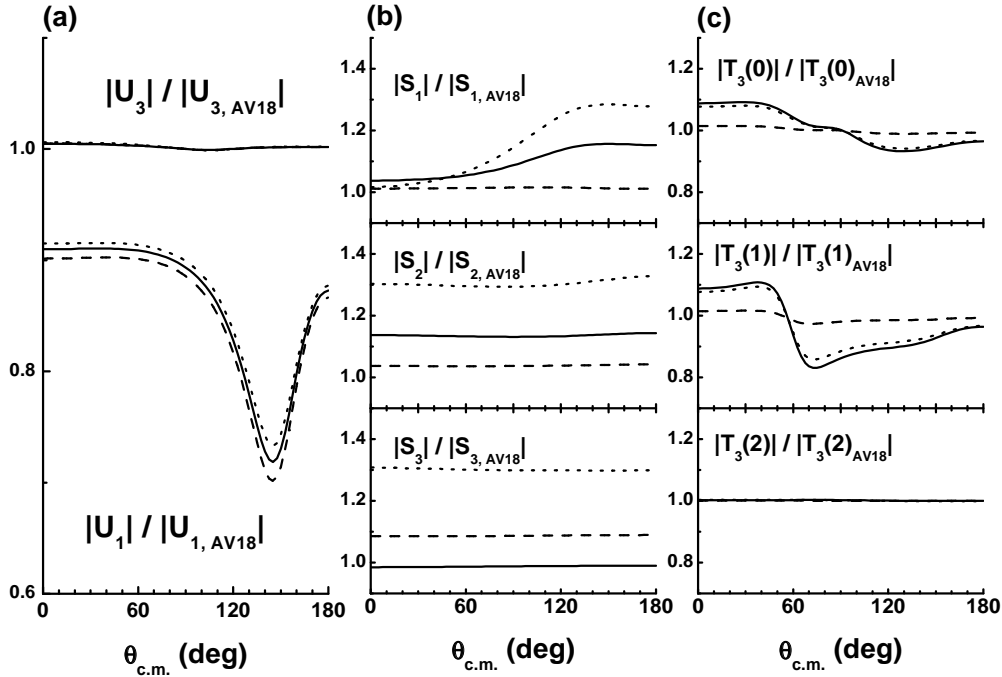}
\caption{\label{fig2}
The 3NF effects on the scattering amplitudes in the $nd$ scattering at $E_n=3$ MeV are shown by the ratio of the magnitude calculated with a 3NF to that without it for 
the scalar amplitudes (a), for the vector amplitudes (b), and for the tensor amplitudes in the spin quartet scattering with $\kappa=0, 1, 2$ (c). 
The solid curves denote the calculations for AV18+BR, the dashed curves for AV18+GS, and the dotted curves for AV18+BR+SO. 
}
\end{figure*}

We first calculate the magnitudes of the scalar, vector and tensor amplitudes, for which the contributions of the 3NF are examined. 
Fig.\ \ref{fig1} shows as functions of the scattering angle the squares of the absolute magnitudes of the amplitudes, $U_1$, $U_3$, $S_1$, $S_2$, $S_3$, $T_1(\kappa)$, $T_2(\kappa)$, and $T_3(\kappa)\; (\kappa=0, 1, 2)$, obtained by the Faddeev calculation without the 3NF for the $nd$ scattering at $E_n = 3$ MeV. 
There the magnitude of the quartet scattering amplitude $U_3$ is much larger than that of the doublet scattering amplitude $U_1$. 
Among the three vector amplitudes, the magnitude of the quartet scattering amplitude $S_3$ is much larger than other two, those of the doublet scattering $S_1$ and the doublet-quartet non-diagonal transition amplitude $S_2$. 
The magnitude of $S_2$ is larger than that of $S_1$. 
They have similar shapes in the angular distribution which have maxima around $\theta=90^{\circ}$. 
The absolute squares of the tensor amplitudes, $T_1(0)$, $T_2(0)$, and $T_3(0)$ show similar angular dependence in a global sense. 
Other tensor amplitudes also have common characteristics in the angular distributions for $\kappa=1$ and $\kappa=2$, respectively, although in less grades than for $\kappa=0$. 
These properties of the vector and tensor amplitudes reflect the specific characters of the $\theta$-dependent factors in Eqs.\ (\ref{eq5}) and (\ref{eq9}), since the $\theta$-dependence of $F(s_i s_fKl_i)$ is weak at the present incident energy.

The contributions of the 3NFs on the $nd$ scattering amplitudes are shown in Fig.\ \ref{fig2}, where the displayed is the ratio of the magnitude of each amplitude calculated with the 3NF to that of the amplitude calculated without any 3NF. 
In the figure, the amplitude $U_3$ is little affected by the 3NFs, while $U_1$ receives large contributions from the BR-3NF as well as from the GS-3NF. 
The SO-3NF provides very small contributions to $U_1$. 
In more detail, the contribution of the BR-3NF to $U_1$ is considered to be mostly due to the central part of the interaction, because the magnitude and the angular distribution of the contribution are very similar to those of the GS-3NF. 
The vector amplitudes, $S_1$, $S_2$, and $S_3$, are influenced by the 3NFs, and particularly the SO-3NF produces large contributions to $S_2$ and $S_3$. 
The 3NF-effects on the tensor amplitudes are examined for example for the spin quartet scattering, where $T_3(0)$ and $T_3(1)$ are considerably affected by the BR-3NF but are very little by the GS-3NF or by the SO-3NF. 
The amplitude $T_3(2)$ is hardly affected by any 3NF studied. 
Such properties of the amplitudes will be reflected on the observables in the analyses in the next section.

\section{Analyses of observables}
\label{Sec_4}

In the previous section, we observed that the scalar amplitudes are larger in magnitude than the vector and tensor amplitudes at the low energy. 
In this section, analytical examinations of $Nd$ observables will be carried in an approximation where we neglect the second order terms of the vector and tensor amplitudes in the observables. 
The results by this approximation, which we call the scalar amplitude dominance (SAD) approximation, are used as the guidelines for the numerical analyses of the  observables with $Nd$ amplitudes provided  by the Faddeev calculations \cite{Is02}.

\subsection{Differential cross section}

The unpolarized differential cross section $\sigma(\theta)$ is expressed as 
\begin{equation}
\sigma(\theta)=\frac16 N_R, 
\label{eq54}
\end{equation}
where  $N_R$ is defined as 
\begin{equation}
N_R=\textrm{Tr}\bigl(\bm{M} \bm{M}^{\dag}\bigr),
\label{eq44}
\end{equation}
and the factor 6 arises from the spin average in the initial state.

In the SAD approximation, $N_R$ is given by
\begin{equation}
N_R=|U_1|^2 +|U_3|^2.
\label{eq47}
\end{equation}

Differences between fully calculated differential cross sections and those in the SAD approximation are less than 1\%.

\subsection{Vector analyzing powers}

\begin{figure*}[t]
\includegraphics[scale=0.5]{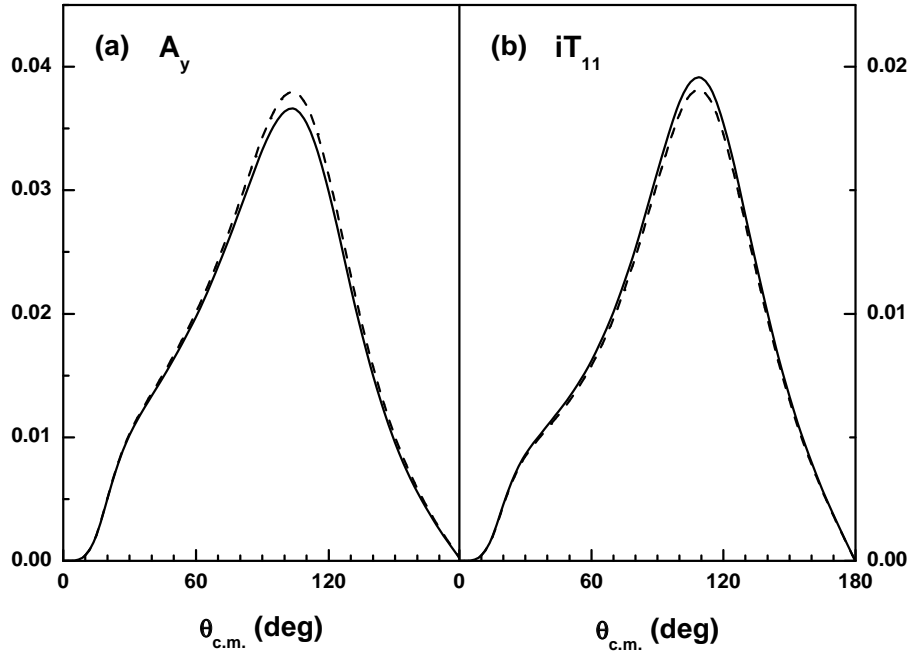}
\caption{\label{fig3}
The vector analyzing powers of protons $A_y$ and deuterons $iT_{11}$ for the $pd$ scattering at $E_p = 3$ MeV by the exact calculations (solid curves) and by the SAD approximation (dashed curves) for AV18.
}
\end{figure*}

The vector analyzing powers of the proton and the deuteron, $A_y$ and $iT_{11}$, are defined as
\begin{eqnarray}
A_y&=&\frac1{N_R} \textrm{Tr}\bigl(\bm{M} {\sigma}_y \bm{M}^{\dag}\bigr),
\\
iT_{11}&=&\frac{i}{N_R} \textrm{Tr}\bigl(\bm{M} \tau_{11} \bm{M}^{\dag}\bigr), 
\label{eq43}
\end{eqnarray}
where ${\tau}_{11}=-\frac{\sqrt3}2(\tau_x +i\tau_y)$ and $\tau_x$ and $\tau_y$ are the $x$ and $y$ components of the spin vector operator of the deuteron.

In the SAD approximation,
\begin{equation}
A_y = \frac4{3N_R} \textrm{Im} \bigl\{ 
  U_1\bigl(-\frac1{\sqrt2} S_1 +2S_2\bigr)^{*} 
 +U_3\bigl(\sqrt2 S_2 +\sqrt{\frac52} S_3\bigr)^{*} \bigr\},
\label{eq45}
\end{equation}
\begin{equation}
iT_{11}= \frac1{N_R} \sqrt{\frac23} \textrm{Im} \bigl\{ 
  U_1\bigl(2S_1-\sqrt2S_2\bigr)^{*}
 +U_3\bigl(-S_2 +\sqrt5 S_3\bigr)^{*} \bigr\}. 
\label{eq46}
\end{equation}
These equations show that the scalar amplitudes and the vector ones dominantly govern the vector analyzing powers.

First we will numerically examine the validity of the SAD approximation in the vector analyzing powers, by comparing Eqs.\ (\ref{eq45}) and (\ref{eq46}) with the exact calculations. 
Fig.\ \ref{fig3} shows the comparison where the SAD approximation works very well, indicating the contributions of the neglected terms to be small.  
Therefore the approximation will have sufficient accuracy to obtain the guidelines for further calculations.

\begin{figure*}[t]
\includegraphics[scale=0.5]{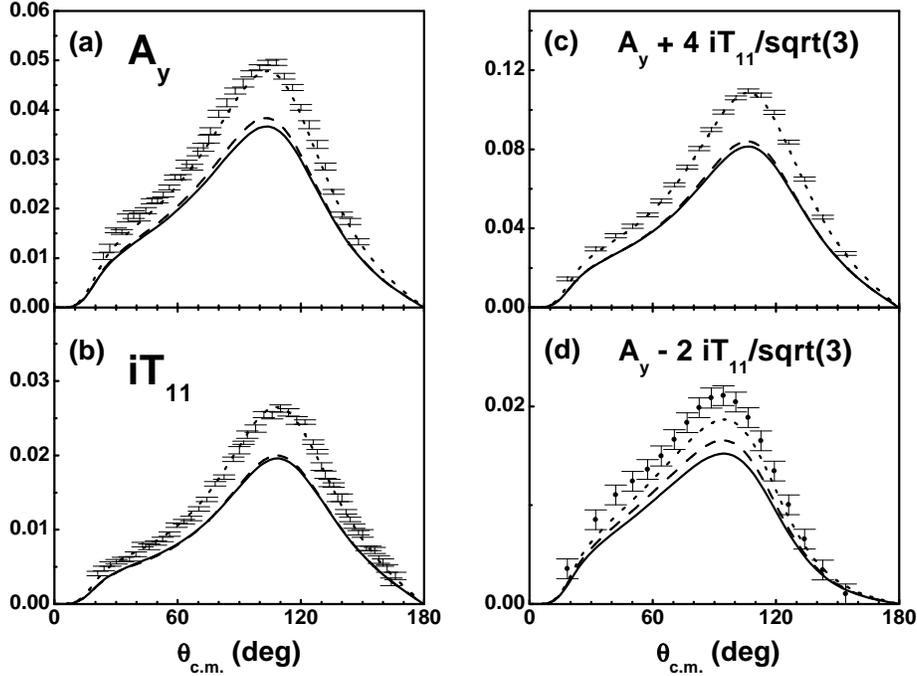}
\caption{\label{fig4}
Comparison of the Faddeev calculations and experimental data \protect\cite{Sh95} for (a) the proton vector analyzing power $A_y$, (b) the deuteron vector analyzing power $iT_{11}$, 
(c) $A_y+\frac4{\sqrt3} iT_{11}$, and 
(d) $A_y-\frac2{\sqrt3} iT_{11}$ in the $pd$ scattering at $E_p = 3$ MeV. 
The dashed curves denote the calculations for AV18, the solid curves for AV18+BR, and the dotted curves for AV18+BR+SO. 
For (c) and (d), quasi-experimental data made by fitting the experimental data of $A_y$ and $iT_{11}$ in Ref.\ \protect\cite{Sh95} are plotted with the error bars.
}
\end{figure*}

Next, we will examine the contributions of the related interactions in detail.  Since the magnitude of $U_3$ is much larger than that of $U_1$ as seen in Fig.\ \ref{fig1}, one can expect the dominant contribution to the vector analyzing powers to arise from the $U_3$-terms of Eqs.\ (\ref{eq45}) and (\ref{eq46}), which are accompanied by two vector amplitudes $S_2$ and $S_3$. 
To extract the contribution of one of these amplitudes, we will eliminate the other by considering a linear combination of $A_y$ and $iT_{11}$: 
the contribution of $S_2$ will be enhanced by the combination $A_y -\frac2{\sqrt3} iT_{11}$ and that of $S_3$ by the combination $A_y+\frac4{\sqrt3} iT_{11}$. 
Considering these features, we will investigate for the $pd$ scattering the contribution of the 2NF and the corrections due to the 3NF to the analyzing powers where the SO-3NF produces the dominant correction. 
As seen in Fig.\ \ref{fig4}, the Faddeev calculation with the AV18 does not reproduce the measured $A_y$ and $iT_{11}$, while the calculation which includes the SO-3NF improves very much the agreement with the data for both of $A_y$ and $iT_{11}$ due to the large contribution of the SO-3NF to the amplitudes $S_2$ and $S_3$ as seen in Fig.\ \ref{fig2}. 
In the figure, the calculations are compared with the data for the combinations $A_y+\frac4{\sqrt3} iT_{11}$ and $A_y-\frac2{\sqrt3} iT_{11}$ obtained from those of $A_y$ and $iT_{11}$, where the SO-3NF reproduces quite well the data of the former combination but the data of the latter are not sufficiently reproduced by the calculation, although the improvement of the agreement with the latter data is appreciable for the SO-3NF contribution. 
This means that the SO-3NF does not describe the amplitude $S_2$ sufficiently. 
Since the magnitude of $S_2$ is much smaller than that of $S_3$ as seen in Fig.\ \ref{fig1}, the inadequacy of the contribution of $S_2$ is masked by the contribution of $S_3$ in $A_y$ and $iT_{11}$. 
Then, the linear combinations proposed here will provide more refined tests of the spin vector interaction than the analyzing powers themselves, at such low energies.  

\subsection{Spin correlation coefficients}

\begin{figure*}[t]
\includegraphics[scale=0.5]{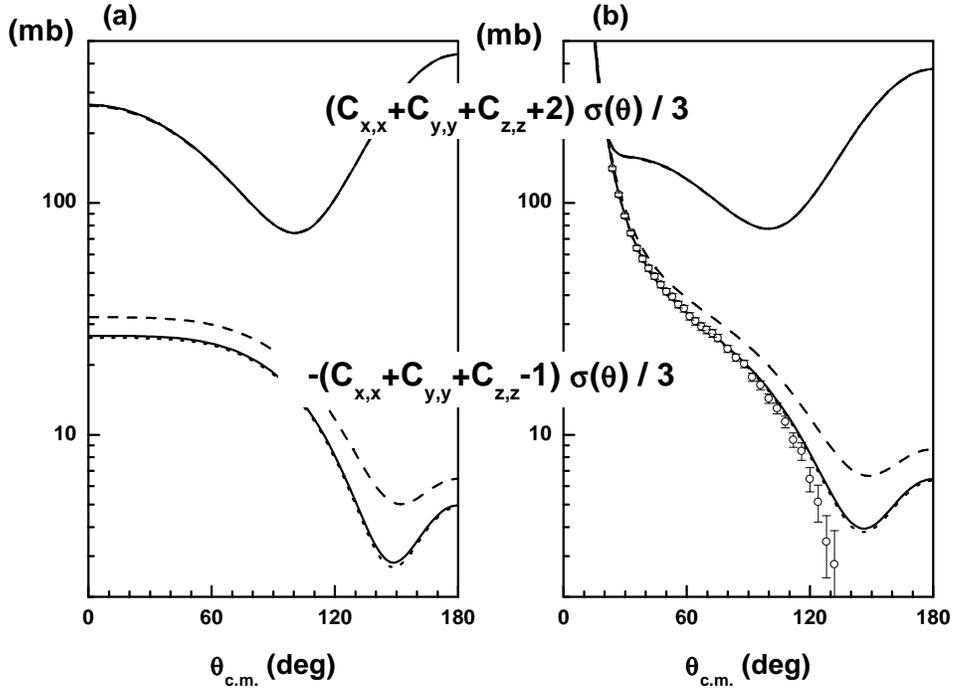}
\caption{\label{fig5}
Effects of 3NFs on spin correlation coefficients for vector polarizations of deuterons. 
The calculated $\bigl(C_{x,x} + C_{y,y} + C_{z,z} + 2 \bigl) \sigma(\theta)/3$ 
and $-\bigl(C_{x,x} +C_{y,y} +C_{z,z} -1\bigl) \sigma(\theta)/3$ 
are shown for the $nd$ and $pd$ scattering at $E_N=3$ MeV in (a) and (b), respectively, where the dashed curves denote the calculations for AV18, 
the solid curves for AV18+BR, and the dotted curves for AV18+GS.
See the text for the points with error bars.
}
\end{figure*}

In the previous paper \cite{Is01}, we have investigated the total cross sections of the $nd$ scattering for the unpolarized neutron and deuteron, for the transversal polarizations where the polarizations of the neutron and the deuteron are perpendicular to the $z$ axis, and for the longitudinal polarizations where the polarizations of the neutron and the deuteron are parallel to the $z$ axis, and have shown that linear combinations of these cross sections give information on $U_1(\theta=0)$ and $U_3(\theta=0)$, separately. 
To extend this idea to finite angles, we will investigate the spin correlation coefficients defined as
\begin{equation}
C_{\alpha,\beta} = 
\frac1{N_R} \textrm{Tr}\bigl(\bm{M}{\tau}_\alpha {\sigma}_\beta \bm{M}^{\dag}\bigr). 
\label{eq49}
\end{equation}


The spin correlation coefficient for the transversal polarizations will representatively be described by the average of $C_{x,x}$ and $C_{y,y}$, which is given in the SAD approximation as
\begin{eqnarray}
\frac12 \bigl( C_{x,x} +C_{y,y} \bigr)
&=&\frac1{2N_R} \textrm{Re} \bigl\{ -\frac43 |U_1|^2 +\frac23 |U_3|^2 
\nonumber \\
&&-\frac{2\sqrt2}3 U_1^{*} T_1(0) +\frac23 U_3^{*} T_2(0) 
\nonumber \\
&&-\frac43 U_3^{*} T_3(0) \bigr\}
\label{eq50}
\end{eqnarray}
and the spin correlation coefficient for the longitudinal polarization $C_{z,z}$ in the SAD approximation is given by
\begin{eqnarray}
C_{z,z} &=& \frac1{N_R} \textrm{Re} \bigl\{ -\frac23 |U_1|^2 +\frac13 |U_3|^2 
\nonumber \\
&&+\frac{2\sqrt2}3 U_1^{*} T_1(0) -\frac23 U_3^{*} T_2(0) 
\nonumber \\
&&+\frac43 U_3^{*} T_3(0) \bigr\}.
\label{eq51}
\end{eqnarray}
These results suggest that one can get $-2 |U_1|^2 + |U_3|^2$ in terms of the spin correlation observables by taking a linear combination of Eqs.\ (\ref{eq50}) and (\ref{eq51}). 
Further, by using the unpolarized differential cross section $\sigma(\theta)$, we get
\begin{equation}
|U_1|^2 = - 2 \bigl(C_{x,x} +C_{y,y} +C_{z,z}-1\bigr) \sigma(\theta)
\label{eq52}
\end{equation}
and
\begin{equation}
|U_3|^2= 2  \bigl(C_{x,x} +C_{y,y} +C_{z,z} +2\bigr) \sigma(\theta),
\label{eq53}
\end{equation}
where we have used Eqs.\ (\ref{eq54}) and (\ref{eq44}) with Eq.\ (\ref{eq47}).

For both of the $pd$ and $nd$ scattering, 
$-\frac13 \bigl(C_{x,x} +C_{y,y} +C_{z,z}-1\bigr) \sigma(\theta)$ and 
$\frac13 \bigl(C_{x,x} +C_{y,y} +C_{z,z} +2\bigr) \sigma (\theta)$ 
obtained by the full calculations are displayed in Fig.\ \ref{fig5}. 
These quantities describe $\frac16 |U_1|^2$ and $\frac16 |U_3|^2$, respectively, in the SAD approximation and we see again that $|U_3|^2$ is much larger than $|U_1|^2$ for the $nd$ scattering as in Fig.\ \ref{fig1} and also for the $pd$ scattering except for small angles where the Coulomb interaction dominates. 
The contributions of the BR-3NF and the GS-3NF are displayed in the figure, where one can see that the contributions are remarkable in $|U_1|^2$ but not in $|U_3|^2$ and the GS-3NF is a good simulation of the scalar part of the BR-3NF.  Such 3NF contributions will experimentally be examined by measuring the cross section and the spin correlation coefficients.



In Fig.\ \ref{fig5} (b), we plot the data of approximate doublet cross sections obtained by 
\begin{equation}
\sigma_1(\theta) \equiv \sigma^\textrm{exp}(\theta)-\frac16|U_3^\textrm{cal}|^2,
\end{equation}
where the experimental data in Ref.\ \cite{Sa94} are used for $\sigma^\textrm{exp}(\theta)$ and the results of the AV18+BR calculation for $U_3^\textrm{cal}$.
Since the amplitude $U_3$ is little affected by the 3NFs as observed in Fig. \ref{fig2}, the use of another calculation for $U_3^\textrm{cal}$ produces essentially the same results. 
The theoretical prediction of 
$-\frac13 \bigl(C_{x,x} +C_{y,y} +C_{z,z}-1\bigr) \sigma(\theta)$ 
agrees well to the data $\sigma_1(\theta)$ up to $\theta\sim120^\circ$ when the 3NF is included, showing the 3NF contribution to be indispensable to describe the doublet scattering.

The remarkable effect of the 3NF on $U_1$ is related to the effect on the triton binding energy, as will be understood by the characteristic that the doublet scattering amplitudes at low energies are governed by a position of the 3N bound state pole \cite{Is99}. 
Then it will be important to confirm such theoretical predictions of the characteristics of the $nd$ central interaction by the experimental measurements to fully understand the role of the scalar part of the 2$\pi$E-3NF at low energies. 
Another combinations of $C_{x,x}$, $C_{y,y}$, and $C_{z,z}$ will be discussed later together with tensor analyzing powers.

\begin{figure*}[t]
\includegraphics[scale=0.5]{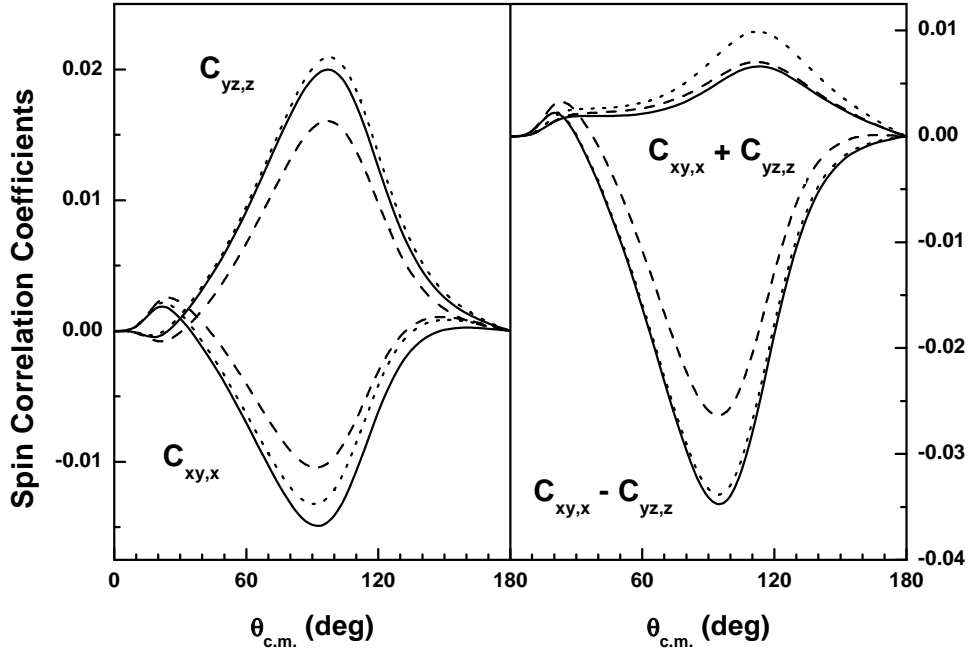}
\caption{\label{fig6}
Effects of 3NFs on the spin correlation coefficients $C_{xy,x}$ and $C_{yz,z}$, and their linear combinations $C_{xy,x} +C_{yz,z}$ and $C_{xy,x} -C_{yz,z}$ for the $pd$ scattering at $E_p = 3$ MeV.
The dashed curves denote the calculations for AV18, 
the solid curves for AV18+BR, and the dotted curves for AV18+BR+SO.
}
\end{figure*}

Further, we will examine $C_{x,z}$ and $C_{z,x}$ as other examples of the spin correlation coefficients for the vector polarizations of the nucleon and the deuteron. 
Generally, the contributions of the vector amplitudes and those of the tensor ones will be mixed up in these coefficients. 
However, their effects are separated to each other by considering their linear combinations.  
In fact, in the SAD approximation,
\begin{eqnarray}
C_{x,z} +C_{z,x} &=& \frac2{{\sqrt3} N_R} \textrm{Re} \bigl\{ -2U_1^{*} T_1(1) 
\nonumber \\
&& +\sqrt2 U_3^{*} \bigl( T_2(1)-2T_3(1) \bigr) \bigr\},
\label{eq572}
\end{eqnarray}
\begin{equation}
C_{x,z}-C_{z,x}=-\frac{2\sqrt2}{N_R} 
  \textrm{Re} \bigl\{ \bigl(\sqrt2 U_1 -U_3\bigr)^{*} S_2 \bigr\}.
\label{eq573}
\end{equation}
The combination $C_{x,z}+C_{z,x}$  consists of the scalar amplitudes and the tensor ones and will reflect the contributions of the central interactions and the tensor ones. 
On the contrary, the combination $C_{x,z}-C_{z,x}$ will describe the contributions of the central interactions and the spin vector interactions. 
The former quantity will be investigated by considering further combinations with tensor analyzing powers. 
The latter quantity will exhibit the contribution of the SO-3NF to the vector amplitude $S_2$, since $U_1$ is small and $U_3$ is insensitive to the 3NF.

Next we will investigate $C_{xy,x}$ and $C_{yz,z}$ as examples of the spin correlation coefficients due to the tensor polarizations of the deuteron.
In the SAD approximation,
\begin{eqnarray}
C_{xy,x} +C_{yz,z} &=& \frac32 C_{yy,y} 
\nonumber\\ 
 &=& \frac2{N_R} \textrm{Im} \bigl\{ -U_1(2\sqrt2 S_1 +S_2)^{*}
\nonumber\\
 &&+U_3(-\frac1{\sqrt2} S_2+ \sqrt{\frac25} S_3)^{*} \bigr\},
\label{eq56}
\end{eqnarray}
\begin{eqnarray}
C_{xy,x}-C_{yz,z} &=& -\bigl(C_{xx,y}+\frac12 C_{yy,y}\bigr) 
\nonumber\\ 
&=& \frac{2\sqrt3}{N_R} \textrm{Im} 
  \bigl\{ U_1 T_1(1)^{*} +\frac1{\sqrt2} U_3 T_2(1)^{*} \bigr\}.
\nonumber \\
~
\label{eq57}
\end{eqnarray}
Due to Eqs.\ (\ref{eq56}) and (\ref{eq57}), $C_{xy,x} +C_{yz,z}$ characterizes the contribution of the vector amplitudes and $C_{xy,x}-C_{yz,z}$ that of the tensor ones. 
Further it is expected that the SO-3NF contributes to the former quantity through the vector amplitudes $S_2$ and $S_3$ and the BR-3NF to the latter through the tensor amplitude $T_2(1)$, since the magnitude of $U_1$ is small compared to that of $U_3$ and $U_3$ is hardly affected by the 3NF.

In Fig.\ \ref{fig6}, the calculated $C_{xy,x}$ and $C_{yz,z}$ and their linear combinations, $C_{xy,x} +C_{yz,z}$ and $C_{xy,x}-C_{yz,z}$ for the $pd$ scattering are displayed for the three kinds of interactions, AV18, AV18+BR, and AV18+BR+SO. 
In the calculated $C_{xy,x}$ and $C_{yz,z}$, the vector effect of the SO-3NF and the tensor effect of the BR-3NF are actually mixed up as seen in the left panel of Fig.\ \ref{fig6}.
However, in the right panel of Fig.\ \ref{fig6}, they are clearly separated to each other, that is, the effect of the SO-3NF appears in $C_{xy,x} +C_{yz,z}$ but not in $C_{xy,x}-C_{yz,z}$, while the effect of the BR-3NF appears in the latter combination but not in the former one. 
These characteristics are consistent with the theoretical prediction by Eqs.\ (\ref{eq56}) and (\ref{eq57}). 
Therefore, measurements of these spin correlation coefficients will be useful to identify the contributions of these 3NF, separately.  

\subsection{Tensor analyzing powers of deuterons}

Tensor analyzing powers of the deuteron $T_{2\kappa}\; (\kappa=0, 1, 2)$ are defined by
\begin{equation}
T_{2\kappa}=\frac1{N_R} \textrm{Tr}
  \bigl(\bm{M} {\tau}_{2\kappa} {\bm{M}}^{\dag}\bigr),
\label{eq58}
\end{equation}
where ${\tau}_{2\kappa}$ is the spin tensor operator of the deuteron with the $z$ component of $\kappa$. 
In the SAD approximation,
\begin{equation}
T_{2\kappa} = \frac1{N_R} \textrm{Re} \bigl\{ -2U_1^{*} T_1(\kappa) 
  +\sqrt2 U_3^{*} \bigl(T_2(\kappa) +T_3(\kappa)\bigr) \bigr\}.
\label{eq59}
\end{equation}

Then, the tensor analyzing powers represent the contributions of the scalar amplitudes and the tensor ones. 
On the other hand, the tensor amplitudes are influenced by the tensor part of the BR-3NF, as shown in Fig. \ref{fig2}, where the magnitudes of $T_3(0)$ and $T_3(1)$ are affected by the 3NF at most angles, while the magnitude of $T_3(2)$ is not. 
 Such 3NF effects can be extracted by the linear combination of the tensor analyzing powers and the spin correlation coefficients. 
Using Eqs.\ (\ref{eq50}) and (\ref{eq51}), we get in the SAD approximation,
\begin{eqnarray}
&&  \frac1{\sqrt2}  \bigl(C_{x,x} +C_{y,y}\bigr) -\sqrt2 C_{z,z}
\nonumber \\
&=&   \frac1{N_R} \textrm{Re} \bigl\{-2U_1^{*} T_1(0) 
   +\sqrt2 U_3^{*} \bigl(T_2(0)-2T_3(0)\bigr) \bigr\}.
\nonumber \\
 ~
\label{eq60}
\end{eqnarray}
From Eqs.\ (\ref{eq59}) and (\ref{eq60}), we obtain
\begin{equation}
T_{20} -\frac1{\sqrt2} \bigl(C_{x,x} +C_{y,y}\bigr) +\sqrt2 C_{z,z} =
  \frac{3\sqrt2}{N_R} \textrm{Re}\bigl(U_3^{*} T_3(0)\bigr).
\label{eq61}
\end{equation}
Relations similar to Eq.\ (\ref{eq61}) are derived for the tensor amplitudes $T_3(1)$ and $T_3(2)$ in the SAD approximation,
\begin{equation}
T_{21} -\frac{\sqrt3}2 \bigl(C_{x,z}+C_{z,x}\bigr)
  =\frac{3\sqrt2}{N_R} \textrm{Re}\bigl(U_3^{*}T_3(1)\bigr),
\label{eq574} 
\end{equation}
\begin{equation}
T_{22}+\frac{\sqrt3}2 \bigl(C_{x,x}-C_{y,y}\bigr)=
  \frac{3\sqrt2}{N_R}  \textrm{Re}\bigl(U_3^{*}T_3(2)\bigr).
\label{eq62}
\end{equation}

\begin{figure}[t]
\includegraphics[scale=0.4]{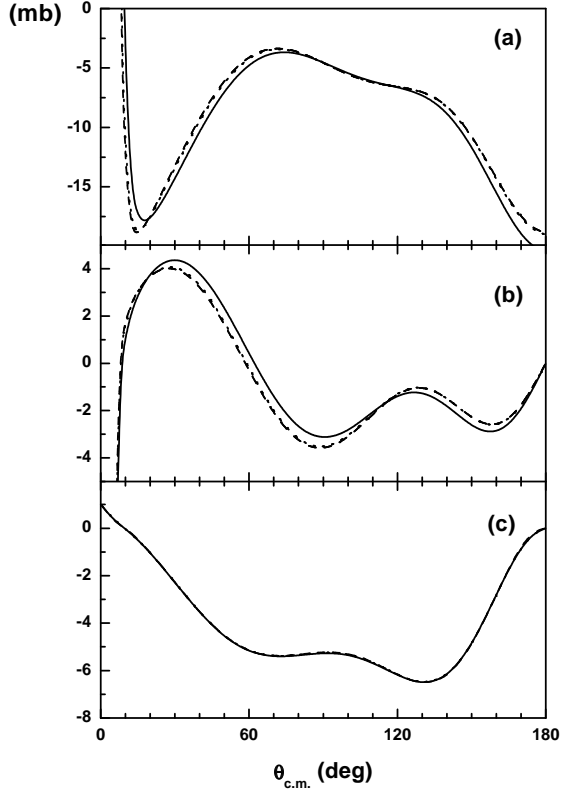}
\caption{\label{fig7}
Effects of 3NFs on tensor amplitudes in the $pd$ scattering at $E_p = 3$ MeV. 
The quantities, 
$\Bigl(T_{20}- \frac1{\sqrt2} \bigl(C_{x,x} +C_{y,y}\bigr) +\sqrt2 C_{z,z}\Bigr)\sigma(\theta)$, 
$\Bigl(T_{21}-\frac{\sqrt3}2 \bigl(C_{x,z}+C_{z,x}\bigr)\Bigr)\sigma(\theta)$, 
and 
$\Bigl(T_{22}+\frac{\sqrt3}2\bigl(C_{x,x}-C_{y,y}\bigr)\Bigr)\sigma(\theta)$ 
(see text) are shown in (a), (b), and (c), respectively. 
The dashed curves denote the calculations for AV18, the solid curves for AV18+BR, and the dotted curves for AV18+GS. 
The dashed curves and the dotted curves are overlapped almost completely in (a) and (b), and the three kinds of curves cannot be identified to each other in (c) .
}
\end{figure}

Since $U_3$ is insensitive to the 3NF, the right hand sides of Eqs.\ (\ref{eq61}), (\ref{eq574}), (\ref{eq62}) will reflect the tensor effect of the BR-3NF on $T_3(0)$, $T_3(1)$ and $T_3(2)$ shown in Fig. \ref{fig2}. 
To eliminate the 3NF effect on $N_R$, we will show the calculated 
$\Bigl(T_{20}- \frac1{\sqrt2} \bigl(C_{x,x} +C_{y,y}\bigr) +\sqrt2 C_{z,z}\Bigr)\sigma(\theta)$, 
$\Bigl(T_{21}-\frac{\sqrt3}2 \bigl(C_{x,z}+C_{z,x}\bigr)\Bigr)\sigma(\theta)$, 
and 
$\Bigl(T_{22}+\frac{\sqrt3}2\bigl(C_{x,x}-C_{y,y}\bigr)\Bigr)\sigma(\theta)$ in Fig.\ \ref{fig7}. 
The figure shows the 3NF tensor effects to be consistent with the characteristics of the effects on tensor amplitudes in Fig.\ \ref{fig2}.  
That is, the 3NF tensor effect is small but finite in $\textrm{Re}\bigl(U_3^{*}T_3(0)\bigr)$ and $\textrm{Re}\bigl(U_3^{*}T_3(1)\bigr)$, while the effect is almost negligible in $\textrm{Re}\bigl(U_3^{*}T_3(2)\bigr)$. 
These features can be examined by measuring the cross section and the spin correlation coefficients and such measurements will be important to determine the contribution of the 2NF tensor interactions and the 3NF tensor effect in the scattering, since $T_3(0)$, $T_3(1)$ and $T_3(2)$ form a complete set of the tensor amplitude in the spin quartet scattering.   

\section{Concluding remarks}
\label{Sec_5}

We have proposed some combinations of the scattering observables for obtaining the detailed information on the contributions of the 2NF and various models of the 3NF.  
Since each combination characterizes the contribution of a particular interaction, measurements of these quantities will provide clear tests for the validity of the interaction.

The numerical investigations are particularly focused on the contributions of the 2$\pi$E-3NF and the SO-3NF. 
As is well-known the 2$\pi$E-3NF provides the indispensable contribution to the triton binding energy, and by the present analyses it turns out that this 3NF effect produces clear contributions to the spin doublet scalar amplitude in the $Nd$ scattering, which can be examined by measuring the cross section and some spin correlation coefficients. 
Further, the Faddeev calculations clarify that the 2$\pi$E-3NF contributes also as the tensor interaction to the tensor amplitudes in the spin quartet scattering. 
These 3NF effects will be criticized by comparing with experimental data when the related observables are measured. 
The SO-3NF produces the remarkable contribution to the vector amplitudes in the quartet scattering and also in the doublet-quartet nondiagonal transition. 
This improves the calculated vector analyzing powers of the proton and the deuteron successfully, although the agreement with the experimental data transformed for the nondiagonal transition is not so good as that for the spin quartet scattering.

The SAD approximation is useful to find the suitable combination of the spin observables for the examination of a particular interaction. 
In the present investigation, the analyzing powers and the spin correlation coefficients are chosen as the spin observables.
However, polarization transfer coefficients will also be available for the examination of the interaction. 
For the convenience of such applications, we will give in Appendix \ref{App_B} the formulas of the polarization transfer coefficients in the SAD approximation, which will be useful in finding of suitable combinations for the test of the validity of particular interactions.

\begin{acknowledgments}
This research was supported by the Japan Society for the Promotion of Science, 
under Grants-in-Aid for Scientific Research No. 13640300.
The numerical calculations was supported, in part, by the Computational Science 
Research Center, Hosei University, under Project No. lab0003.
\end{acknowledgments}

\appendix
\section{Matrix elements of $\bm{M}$ in terms of the invariant amplitudes}
\label{App_A}

The matrix elements $A$, ..., $R$ in Eq.\ (\ref{eq1}) are described by the invariant amplitudes $U_1$, ..., $V(\kappa)$ as follows:
\begin{equation}
A = \frac12 \bigl(U_3+ T_3(0)\bigr)
\label{eq16}
\end{equation}
\begin{equation}
B=\frac1{\sqrt6} \bigl(T_2(1) -T_3(1)\bigr) -\frac1{\sqrt2} S_4 
  +\frac1{\sqrt{10}} S_3 +\frac1{\sqrt{15}} V(1)
\label{eq17}
\end{equation} 
\begin{equation}
C=-\frac1{\sqrt3} \bigl(\frac12 T_2(1) +T_3(1)\bigr) 
  +\frac12 S_4 +\frac1{\sqrt5} S_3 +\sqrt{\frac2{15}} V(1)
\label{eq18}
\end{equation}
\begin{equation}
D=\frac1{\sqrt3} \bigl(-T_2(2)+T_3(2) - V(2)\bigr)
\end{equation}
\begin{equation}
E=\frac1{\sqrt6} \bigl(2T_2(2)+T_3(2) - V(2)\bigr)
\end{equation}
\begin{equation}
F=V(3)
\label{eq19}
\end{equation}
\begin{equation}
G=-\frac1{\sqrt6} \bigl(T_1(1)+T_3(1)\bigr) -\frac1{\sqrt2} S_2 
  -\frac1{\sqrt{10}} S_3 - \frac1{\sqrt{15}} V(1)
\label{eq20}
\end{equation}
\begin{equation}
H=\frac{\sqrt2}3 U_1+\frac16 U_3+\frac16 \bigl(-2T_1(0)+2T_2(0) -T_3(0)\bigr)
\label{eq21}
\end{equation}
\begin{equation}
I=-\frac13 U_1 +\frac1{3\sqrt2} U_3 
  -\frac1{3\sqrt2} \bigl(2T_1(0) +T_2(0) +T_3(0)\bigr)
\label{eq22}
\end{equation}
\begin{eqnarray}
J&=&\frac1{2\sqrt3} \bigl(2T_1(1) -T_2(1)\bigr) 
  +\frac{\sqrt2}3 S_1 -\frac16 S_4 -\frac13 S_2 
\nonumber\\
 && +\frac2{3\sqrt5} S_3 -\sqrt{\frac2{15}} V(1)
\label{eq23}
\end{eqnarray}
\begin{eqnarray}
K&=&\frac1{\sqrt6} \bigl(T_1(1) +T_2(1)\bigr) -\frac23 S_1 
  +\frac1{3\sqrt2} \bigl(S_4 -S_2\bigr) 
\nonumber\\
  &&+\frac{\sqrt2}{3\sqrt5} S_3 
  -\frac1{\sqrt{15}} V(1)
\label{eq24}
\end{eqnarray}
\begin{equation}
L=\frac1{\sqrt6}\bigl(-2T_1(2) +T_3(2) +V(2)\bigr)
\label{eq25}
\end{equation}
\begin{equation}
M=\frac1{2\sqrt3} \bigl(T_1(1) -2T_3(1)\bigr) 
   +\frac12 S_2 -\frac1{\sqrt5} S_3 -\sqrt{\frac2{15}} V(1)
\label{eq26}
\end{equation}
\begin{equation}
N=\frac1{3\sqrt2} \bigl(-\sqrt2 U_1 +U_3 + T_1(0) +2T_2(0)-T_3(0)\bigr)
\label{eq27}
\end{equation}
\begin{equation}
O=\frac13\bigl(\frac1{\sqrt2} U_1 +U_3 +T_1(0) -T_2(0) -T_3(0)\bigr)
\label{eq28}
\end{equation}
\begin{eqnarray}
P&=&-\frac1{\sqrt{6}} \bigl(T_1(1) +T_2(1)\bigr) -\frac13 S_1 
+\frac1{3\sqrt2} \bigl(-S_4 +S_2\bigr)
\nonumber\\
 &&  +\frac{2\sqrt2}{3\sqrt5} S_3 
  -\frac2{\sqrt{15}} V(1)
\label{eq29}
\end{eqnarray}
\begin{eqnarray}
Q&=&\frac1{2\sqrt3} \bigl(-T_1(1) +2T_2(1)\bigr) 
  +\frac16 \bigl(2\sqrt2 S_1 +2S_4 +S_2\bigr) 
\nonumber\\
 &&+\frac2{3\sqrt5} S_3 
  -\sqrt{\frac2{15}} V(1)
\label{eq30}
\end{eqnarray}
\begin{equation}
R=\frac1{\sqrt3}\bigl(T_1(2) +T_3(2)\bigr) +\frac1{\sqrt3} V(2)
\label{eq31}
\end{equation}

\section{Polarization transfer coefficients in the SAD approximation}
\label{App_B}

\subsection{Deuteron (vector) to Nucleon transfers}

The polarization transfer coefficient is defined as 
\begin{equation}
K_{\alpha}^{\beta}(dn)=\frac1{N_R} \textrm{Tr} \bigl(\bm{M} \tau_{\alpha} 
   {\bm{M}}^{\dag} \sigma_{\beta}\bigr).
\label{eqC1}
\end{equation}

Define $X_1$, $Y_1(\kappa=0, 1, 2)$, and $Z_1$ by
\begin{equation}
X_1 \equiv 
 - \frac23|U_1|^2 + \frac53|U_3|^2
    - \frac{4\sqrt2}3 \textrm{Re}\bigl( U_1 U_3^{*}\bigr),
\label{eqC2}
\end{equation}
\begin{eqnarray}
Y_1(\kappa) &\equiv& \frac{\sqrt2}3 \textrm{Re} 
  \bigl\{U_1\bigl(T_1(\kappa) -4T_2(\kappa) +2T_3(\kappa)\bigr)^{*} 
\nonumber\\
&&  +\sqrt2 U_3\bigl(T_1(\kappa) +\frac12 T_2(\kappa)
   +2T_3(\kappa)\bigr)^{*} \bigr\},
\label{eqC3}
\end{eqnarray}
\begin{equation}
Z_1 \equiv \textrm{Re} \bigl\{ \bigl(\sqrt2 U_1 -U_3\bigr) 
   \bigl(S_1 +\frac5{2\sqrt2} S_2 +\sqrt{\frac52} S_3\bigr)^{*} \bigr\}.
\label{eqC4}
\end{equation}

Then we get
\begin{equation}
K_x^x(dn) +K_y^y(dn)=\frac2{3N_R} \bigl(X_1-Y_1(0)\bigr),
\label{C5}
\end{equation}
\begin{equation}
K_z^z(dn)=\frac1{3N_R} \bigl(X_1+2Y_1(0)\bigr),
\label{eqC6}
\end{equation}
\begin{equation}
K_x^x(dn)-K_y^y(dn)=\frac{2\sqrt2}{\sqrt3 N_R} Y_1(2),
\label{eqC7}
\end{equation}
\begin{equation}
K_z^x(dn) + K_x^z(dn) = -\frac{2\sqrt2}{\sqrt3 N_R} Y_1(1),
\label{eqC8}
\end{equation}
\begin{equation}
K_z^x(dn)-K_x^z(dn)=\frac8{9N_R} Z_1.
\label{eqC9}
\end{equation}

\subsection{Nucleon to Deuteron (vector) transfers}

The polarization transfer coefficient is defined as
\begin{equation}
K_{\alpha}^{\beta} (nd)=\frac1{N_R} \textrm{Tr}\bigl(\bm{M} \sigma_{\alpha} 
  {\bm{M}}^{\dag} \tau_{\beta}\bigr).
\label{eqC10}
\end{equation}

Define $X_2$, $Y_2(\kappa=0, 1, 2)$, and $Z_2$ as
\begin{equation}
X_2 \equiv X_1, 
\label{eqC11}
\end{equation}
\begin{eqnarray}
Y_2(\kappa) &\equiv& \frac{\sqrt2}3 \textrm{Re} \bigl\{
  U_1\bigl(4T_1(\kappa)-T_2(\kappa)+2T_3(\kappa)\bigr)^{*} 
\nonumber\\
&&  +\sqrt2 U_3(-\frac12 T_1(\kappa)-T_2(\kappa)+2T_3(\kappa)\bigr)^{*} \bigr\},
\nonumber \\
~~
\label{eqC12}
\end{eqnarray}
\begin{equation}
Z_2 \equiv Z_1.
\label{eqC13}
\end{equation}

Then we get
\begin{equation}
K_x^x(nd) +K_y^y(nd)=\frac2{3N_R} \bigl(X_2-Y_2(0)\bigr),
\label{eqC14}
\end{equation}
\begin{equation}
K_z^z(nd)=\frac1{3N_R}\bigl(X_2+2Y_2(0)\bigr),
\label{eqC15}
\end{equation}
\begin{equation}
K_x^x(nd)-K_y^y(nd)=\frac{2\sqrt2}{\sqrt3N_R}Y_2(2),
\label{eqC16}
\end{equation}
\begin{equation}
K_z^x(nd)+K_x^z(nd)=-\frac{2\sqrt2}{\sqrt3 N_R} Y_2(1),
\label{eqC17}
\end{equation}
\begin{equation}
K_z^x(nd)-K_x^z(nd)=\frac8{9N_R} Z_2.
\label{eqC18}
\end{equation}

\subsection{Deuteron (tensor) to Nucleon transfers}

The polarization transfer coefficient is defined as
\begin{equation}
K_{\alpha \beta}^{\gamma} (dn)=\frac1{N_R} \textrm{Tr}\bigl(\bm{M} 
 \tau_{\alpha \beta} {\bm{M}}^{\dag} \sigma_{\gamma}\bigr).
\label{eqC19}
\end{equation}

Define $X_3$, $Y_3(\kappa=0, 1, 2)$, and $Z_3$ as
\begin{equation}
X_3 \equiv 0,
\label{eqC20}
\end{equation}
\begin{eqnarray}
Y_3(\kappa) &\equiv& \frac2{\sqrt3} \textrm{Im} \bigl\{ 
  U_1\bigl(T_1(\kappa)-2T_3(\kappa)\bigr)^{*} 
\nonumber\\
&&+\sqrt2 U_3\bigl(T_1(\kappa)+\frac32T_2(\kappa)+T_3(\kappa)\bigr)^{*} \bigr\},
\label{eqC21}
\end{eqnarray}
\begin{equation}
Z_3 \equiv \textrm{Im} \bigl\{U_1\bigl(S_2 -\frac2{\sqrt5}S_3\bigr)^{*}
-2U_3\bigl(S_1+\frac3{2\sqrt2}S_2-\sqrt{\frac25} S_3\bigr)^{*} \bigr\}.
\label{eqC22}
\end{equation}

Then we get
\begin{equation}
K_{xx}^y(dn)-K_{yy}^y(dn)=\frac1{N_R} \bigl(-Y_3(1)-2Z_3\bigr),
\label{eqC23}
\end{equation}
\begin{equation}
K_{xx}^y(dn)+K_{yy}^y(dn)=\frac1{N_R} \bigl(-Y_3(1)+\frac23 Z_3\bigr),
\label{C24}
\end{equation}
\begin{equation}
K_{xy}^z(dn)=\frac1{N_R} Y_3(2),
\label{eqC25}
\end{equation}
\begin{equation}
K_{yz}^z(dn)=\frac1{N_R} \bigl(-\frac12 Y_3(1) +Z_3\bigr),
\label{eqC26}
\end{equation}
\begin{equation}
K_{yz}^x(dn)-K_{xz}^y(dn)=-\frac1{N_R}\sqrt{\frac32}Y_3(0).
\label{eqC27}
\end{equation}

\subsection{Nucleon to Deuteron (tensor) transfers}

The polarization transfer coefficient is defined as
\begin{equation}
K_{\alpha}^{\beta \gamma}(nd)=\frac1{N_R} \textrm{Tr}\bigl(\bm{M} 
 \sigma_{\alpha} {\bm{M}}^{\dag} \tau_{\beta \gamma}\bigr).
\label{eqC28}
\end{equation}

Define $X_4$, $Y_4(\kappa=0, 1, 2)$, and $Z_4$ as
\begin{equation}
X_4 \equiv 0,
\label{eqC29}
\end{equation}
\begin{eqnarray}
Y_4(\kappa) &\equiv& \frac2{\sqrt3} \textrm{Im} \bigl\{ 
  U_1\bigl(T_2(\kappa) +2T_3(\kappa)\bigr)^{*} 
\nonumber\\
 &&+\sqrt2 U_3\bigl(\frac32 T_1(\kappa) +T_2(\kappa) -T_3(\kappa)\bigr)^{*} \bigr\},
\label{eqC30}
\end{eqnarray}
\begin{equation}
Z_4 \equiv Z_3.
\label{eqC31}
\end{equation}

Then we get
\begin{equation}
K_y^{xx}(nd) -K_y^{yy}(nd)=\frac1{N_R} \bigl(-Y_4(1)-2Z_4\bigr),
\label{eqC32}
\end{equation}
\begin{equation}
K_y^{xx}(nd) +K_y^{yy}(nd)=\frac1{N_R}\bigl(-Y_4(1)+\frac23 Z_4\bigr),
\label{C33}
\end{equation}
\begin{equation}
K_z^{xy}(nd)=\frac1{N_R} Y_4(2),
\label{C34}
\end{equation}
\begin{equation}
K_z^{yz}(nd)=\frac1{N_R} \bigl(-\frac12 Y_4(1) +Z_4\bigr),
\label{eqC35}
\end{equation}
\begin{equation}
K_x^{yz}(nd) +K_y^{xz}(nd)=-K_z^{xy}(nd),
\label{eqC36}
\end{equation}
\begin{equation}
K_x^{yz}(nd)-K_y^{xz}(nd)=-\frac1{N_R}\sqrt{\frac32} Y_4(0).
\label{eqC37}
\end{equation}

\subsection{Deuteron (vector) to Deuteron (vector) transfers}

The polarization transfer coefficient is defined as
\begin{equation}
K_{\alpha}^{\beta}(dd)= \frac1{N_R} \textrm{Tr}\bigl(\bm{M} \tau_{\alpha} 
  {\bm{M}}^{\dag} \tau_{\beta}\bigr).
\label{eqC38}
\end{equation}

Define $X_5$, $Y_5(\kappa=0, 1, 2)$, and $Z_5$ as
\begin{equation}
X_5 \equiv \frac43|U_1|^2 + \frac53 |U_3|^2
 +  \frac{2\sqrt2}3 \textrm{Re}\bigl(U_1 U_3^{*}\bigr),
\label{eqC39}
\end{equation}
\begin{eqnarray}
Y_5(\kappa) &\equiv& \frac43 \textrm{Re} \bigl\{ 
 \sqrt2 U_1\bigl(T_1(\kappa)- T_2(\kappa) +\frac12T_3(\kappa)\bigr)^{*} 
\nonumber\\
&&+\frac12U_3\bigl(T_1(\kappa)-T_2(\kappa)-4T_3(\kappa)\bigr)^{*} \bigr\},
\label{eqC40}
\end{eqnarray}
\begin{eqnarray}
Z_5 &\equiv& \frac{4\sqrt2}9 \textrm{Re} \bigl\{ 
 U_1\bigl(4S_1 +2\sqrt2 S_2 +\sqrt{\frac52}S_3\bigr)^{*} 
\nonumber\\
&& -U_3(\frac1{\sqrt2} S_1 +5S_2 -\sqrt5 S_3\bigr)^{*} \bigr\}.
\label{eqC41}
\end{eqnarray}

Then we get
\begin{equation}
K_x^x(dd) +K_y^y(dd)=\frac1{N_R} \bigl(\frac23 X_5 +\frac13 Y_5(0)\bigr),
\label{eqC42}
\end{equation}
\begin{equation}
K_x^x(dd)-K_y^y(dd)=-\frac1{N_R} \sqrt{\frac23} Y_5(2),
\label{eqC43}
\end{equation}
\begin{equation}
K_x^z(dd) +K_z^x(dd)=\frac1{N_R} \sqrt{\frac23} Y_5(1),
\label{eqC44}
\end{equation}
\begin{equation}
K_x^z(dd)-K_z^x(dd)=\frac1{N_R} Z_5,
\label{eqC45}
\end{equation}
\begin{equation}
K_z^z(dd)=\frac1{3N_R} \bigl( X_5-Y_5(0)\bigr).
\label{eqC46}
\end{equation}

\subsection{Deuteron (vector) to Deuteron (tensor) transfers}

The polarization transfer coefficient is defined as
\begin{equation}
K_{\alpha}^{\beta \gamma}(dd)=\frac1{N_R} \textrm{Tr}\bigl(\bm{M} 
  \tau_{\alpha} {\bm{M}}^{\dag} \tau_{\beta \gamma}\bigr).
\label{eqC47}
\end{equation}

Define $X_6$, $Y_6(\kappa=0, 1, 2)$, and $Z_6$ as
\begin{equation}
X_6 \equiv 0,
\label{eqC48}
\end{equation}
\begin{eqnarray}
Y_6(\kappa) &\equiv& \sqrt{\frac23} \textrm{Im} \bigl\{ 
  \sqrt2 U_1\bigl(2T_2(\kappa) +T_3(\kappa)\bigr)^{*}
\nonumber\\
&& +U_3\bigl(-3T_1(\kappa)+T_2(\kappa)+2T_3(\kappa)\bigr)^{*} \bigr\},
\label{eqC49}
\end{eqnarray}
\begin{equation}
Z_6 \equiv \textrm{Im} \bigl\{ U_1\bigl(2S_2 -\frac1{\sqrt5}S_3\bigr)^{*} 
 -U_3\bigl(S_1 +\frac{2\sqrt2}{\sqrt5} S_3\bigr)^{*} \bigr\}.
\label{eqC50}
\end{equation}

Then we get
\begin{equation}
K_y^{xx}(dd)+K_y^{yy}(dd)=\frac1{N_R}\bigl(Y_6(1)-\frac23 Z_6\bigr),
\label{eqC51}
\end{equation}
\begin{equation}
K_y^{xx}(dd)-K_y^{yy}(dd)=\frac1{N_R} \bigl(Y_6(1)+2Z_6\bigr),
\label{eqC52}
\end{equation}
\begin{equation}
K_x^{xy}(dd)=-\frac12 \bigl(K_y^{xx}(dd)-K_y^{yy}(dd)\bigr),
\label{eqC53}
\end{equation}
\begin{equation}
K_y^{xz}(dd) +K_x^{yz}(dd)=\frac1{N_R} Y_6(2),
\label{eqC54}
\end{equation}
\begin{equation}
K_y^{xz}(dd)-K_x^{yz}(dd)=-\frac1{N_R}\sqrt{\frac32} Y_6(0),
\label{eqC55}
\end{equation}
\begin{equation}
K_z^{xy}(dd)=-\frac1{N_R}Y_6(2),
\label{eqC56}
\end{equation}
\begin{equation}
K_z^{yz}(dd)=\frac1{N_R} \bigl(\frac12 Y_6(1)-Z_6 \bigr).
\label{eqC57}
\end{equation}

\subsection{Deuteron (tensor) to Deuteron (vector) transfers}

The polarization transfer coefficient is defined as
\begin{equation}
K_{\alpha \beta}^{\gamma}(dd)=\frac1{N_R} \textrm{Tr}\bigl(\bm{M} 
 \tau_{\alpha \beta} {\bm{M}}^{\dag} \tau_{\gamma}\bigr).
\label{eqC58}
\end{equation}

Define $X_7$, $Y_7(\kappa=0, 1, 2)$, and $Z_7$ as
\begin{equation}
X_7 \equiv 0,
\label{eqC59}
\end{equation}
\begin{eqnarray}
Y_7(\kappa) &\equiv& \sqrt{\frac23} \textrm{Im} \bigl\{ 
 \sqrt2 U_1\bigl(2T_1(\kappa)-T_3(\kappa)\bigr)^{*} 
\nonumber\\
&& + U_3\bigl(T_1(\kappa)-3T_2(\kappa)-2T_3(\kappa)\bigr)^{*} \bigr\},
\label{eqC60}
\end{eqnarray}
\begin{equation}
Z_7 \equiv Z_6.
\label{eqC61}
\end{equation}

Then we get
\begin{equation}
K_{xx}^y(dd)+K_{yy}^y(dd)=\frac1{N_R} \bigl(Y_7(1)-\frac23 Z_7\bigr),
\label{eqC62}
\end{equation}
\begin{equation}
K_{xx}^y(dd)-K_{yy}^y(dd)=\frac1{N_R} \bigl(Y_7(1) +2Z_7\bigr),
\label{eqC63}
\end{equation}
\begin{equation}
K_{xy}^x(dd)=-\frac12\bigl(K_{xx}^y(dd)-K_{yy}^y(dd)\bigr),
\label{eqC64}
\end{equation}
\begin{equation}
K_{xz}^y(dd)+K_{yz}^x(dd)=-K_{xy}^z(dd)=\frac1{N_R} Y_7(2),
\label{eqC65}
\end{equation}
\begin{equation}
K_{xz}^y(dd) -K_{yz}^x(dd)=-\frac1{N_R} \sqrt{\frac32} Y_7(0),
\label{eqC66}
\end{equation}
\begin{equation}
K_{yz}^z(dd)=\frac1{N_R}\bigl(\frac12 Y_7(1)-Z_7\bigr).
\label{eqC67}
\end{equation}

\subsection{Nucleon to Nucleon transfers}

The polarization transfer coefficient is defined as
\begin{equation}
K_{\alpha}^{\beta}(nn)=\frac1{N_R}\textrm{Tr}\bigl(\bm{M} \sigma_{\alpha} 
  {\bm{M}}^{\dag} \sigma_{\beta}\bigr)
\label{eqC68}
\end{equation}

Define $X_8$, $Y_8(\kappa=0, 1, 2)$, and $Z_8$ as
\begin{equation}
X_8 \equiv \frac13|U_1|^2 + \frac53|U_3|^2 
  +\frac{8\sqrt2}3 \textrm{Re}\bigl(U_1U_3^{*}\bigr),
\label{eqC69}
\end{equation}
\begin{equation}
Y_8(\kappa) \equiv \frac43 \textrm{Re} \bigl\{  \bigl(\sqrt2U_1-U_3\bigr)
 \bigl(T_1(\kappa)-T_2(\kappa)+2T_3(\kappa)\bigr)^{*} \},
\label{eqC70}
\end{equation}
\begin{eqnarray}
Z_8 &\equiv& \frac89 \textrm{Re} \bigl\{ 
 U_1\bigl(\sqrt{\frac12} S_1+ 2S_2 +2\sqrt5 S_3\bigr)^{*} 
\nonumber\\
 &&-U_3 \bigl(2S_1-5\sqrt2 S_2-\sqrt{\frac52}S_3\bigr)^{*} \bigr\}.
\label{eqC71}
\end{eqnarray}

Then we get
\begin{equation}
K_x^x(nn)-K_y^y(nn)=-\frac1{N_R}\sqrt{\frac23}Y_8(2),
\label{eqC72}
\end{equation}
\begin{equation}
K_x^x(nn)+K_y^y(nn)=\frac1{N_R} \bigl(\frac23 X_8+\frac13 Y_8(0)\bigr),
\label{eqC73}
\end{equation}
\begin{equation}
K_z^z(nn)=\frac1{3N_R} \bigl(X_8 -Y_8(0)\bigr),
\label{eqC74}
\end{equation}
\begin{equation}
K_x^z(nn) +K_z^x(nn)=\frac1{N_R}\sqrt{\frac23} Y_8(1),
\label{eqC75}
\end{equation}
\begin{equation}
K_x^z(nn)-K_z^x(nn)=\frac1{N_R}Z_8.
\label{eqC76}
\end{equation}

\subsection{Deuteron (tensor) to Deuteron (tensor) transfers}

The polarization transfer coefficient is defined as
\begin{equation}
K_{\alpha \beta}^{\gamma \delta}(dd)=\frac1{N_R} \textrm{Tr} \bigl(\bm{M} 
  \tau_{\alpha \beta} {\bm{M}}^{\dag} \tau_{\gamma \delta}\bigr).
\label{eqC77}
\end{equation}

Define $X_9$, $Y_9(\kappa=0, 1, 2)$, and $Z_9$ as
\begin{equation}
X_9 \equiv \frac12|U_3|^2 +\sqrt2 \textrm{Re} \bigl(U_1 U_3^{*}\bigr),
\label{eqC78}
\end{equation}
\begin{equation}
Y_9(\kappa) \equiv \textrm{Re}\bigl\{\sqrt2 U_1 T_3(\kappa)^{*} 
  -U_3\bigl(T_1(\kappa)-T_2(\kappa) \bigr)^{*} \bigr\},
\label{eqC79}
\end{equation}
\begin{equation}
Z_9 \equiv \textrm{Re}\bigl\{ \frac3{\sqrt5} U_1S_3^{*} 
 +U_3\bigl(S_1-\sqrt2 S_2 +\sqrt{\frac25} S_3\bigr)^{*} \bigr\}.
\label{eqC80}
\end{equation}

Then we get
\begin{eqnarray}
K_{zz}^{zz}(dd) &=&
 \bigl(K_{xx}^{xx}(dd) +K_{xx}^{yy}(dd)\bigr) 
  +\bigl(K_{yy}^{xx}(dd) +K_{yy}^{yy}(dd)\bigr) 
\nonumber\\
&=& \frac2{N_R} \bigl(X_9-Y_9(0)\bigr),
\label{eqC81}
\end{eqnarray}
\begin{eqnarray}
&&-\bigl(K_{xx}^{xx}(dd)+K_{xx}^{yy}(dd)\bigr)
  +\bigl(K_{yy}^{xx}(dd)+K_{yy}^{yy}(dd)\bigr)
\nonumber\\
&=&\frac{2\sqrt6}{N_R} Y_9(2),
\label{eqC82}
\end{eqnarray}
\begin{eqnarray}
&&\bigl(K_{xx}^{xx}(dd)-K_{xx}^{yy}(dd)\bigr) 
 +\bigl(K_{yy}^{xx}(dd)-K_{yy}^{yy}(dd)\bigr)
\nonumber\\
&=&-\frac{2\sqrt6}{N_R} Y_9(2),
\label{eqC83}
\end{eqnarray}
\begin{eqnarray}
&&\bigl(K_{xx}^{xx}(dd)-K_{xx}^{yy}(dd)\bigr)
 -\bigl(K_{yy}^{xx}(dd)-K_{yy}^{yy}(dd)\bigr)
\nonumber\\
&=&\frac6{N_R} (X_9+Y_9(0)),
\label{eqC84}
\end{eqnarray}
\begin{equation}
K_{xy}^{xy}(dd)=\frac3{2N_R} \bigl(X_9+Y_9(0)\bigr),
\label{eqC85}
\end{equation}
\begin{equation}
K_{xz}^{xz}(dd) +K_{yz}^{yz}(dd)=\frac3{N_R} \bigl(X_9-\frac12 Y_9(0)\bigr),
\label{eqC86}
\end{equation}
\begin{equation}
K_{xz}^{xz}(dd)-K_{yz}^{yz}(dd)=\frac3{N_R}\sqrt{\frac32}Y_9(2),
\label{eqC87}
\end{equation}
\begin{equation}
K_{yz}^{xy}(dd)=\frac3{2N_R}\bigl(\sqrt{\frac32}Y_9(1)-Z_9\bigr),
\label{eqC88}
\end{equation}
\begin{equation}
K_{xz}^{xx}(dd)-K_{xz}^{yy}(dd)=\frac3{N_R}\bigl(\sqrt{\frac32}Y_9(1)-Z_9\bigr),
\label{eqC89}
\end{equation}
\begin{equation}
K_{xz}^{xx}(dd)+K_{xz}^{yy}(dd)=-\frac1{N_R} \bigl(\sqrt{\frac32}Y_9(1)+3Z_9\bigr),
\label{eqC90}
\end{equation}
\begin{equation}
K_{xx}^{xz}(dd)+K_{yy}^{xz}(dd)=\frac1{N_R}\bigl(-\sqrt{\frac32}Y_9(1)+3Z_9\bigr),
\label{eqC91}
\end{equation}
\begin{equation}
K_{xx}^{xz}(dd)-K_{yy}^{xz}(dd)=\frac3{N_R}\bigl(\sqrt{\frac32}Y_9(1)+Z_9\bigr),
\label{eqC92}
\end{equation}
\begin{equation}
K_{xy}^{yz}(dd)=\frac12 \bigl(K_{xx}^{xz}(dd)-K_{yy}^{xz}(dd)\bigr).
\label{eqC93}
\end{equation}


\end{document}